# Study of boiling heat transfer on the wall with an improved hybrid lattice Boltzmann model

Anjie Hu, Dong Liu[*]

School of Civil Engineering and Architecture, Southwest University of Science and Technology, Mianyang 621010, PR China

## Abstract

In this paper, the energy non-conservation problem of the hybrid thermal LB phase change model is emphasized. To solve this problem, an infinite volume discrete scheme is introduced to deal with the diffusion term in the energy equation. After verified this model with energy conservation law in the simulation of the one-dimensional heat conduction across the liquid-vapor interface and the liquid film evaporation on the heating wall, the improved hybrid thermal LB model is then applied in the simulation of bubble nucleation and boiling heat transfer. With an innovative nucleation site treatment on the wall, the complete boiling stages as well as the boiling curve were successfully reproduced without adding any heating fluctuations. The bubble nucleation, growth, departure, and coalescence are well captured. Some basic features of boiling heat transfer including bubble nucleation site activation, bubble departure frequency, character of bubble shape at different regime, and horizontal movement of bubble during the boiling process were clearly observed in the numerical results.

---

[*] Corresponding author. Tel.: +86 15281112512

E-mail address: dtld@126.com (Dong L)

# 1. introduction

Boiling heat transfer (BHT) is one of the most effective heat transfer modes. It has been extensively applied in energy and cooling industrial [1, 2] such as nuclear reactors, heavy-vehicle engines, computer chips, micro-electronic devices, and so on. Actually, boiling is an extremely complex and elusive process in which various physical components are involved and interrelated, such as the nucleation, growth, departure, and coalescence of vapor bubbles, the transport of latent heat, and the instability of liquid–vapor interfaces. In the last century, efforts have been put to experimentally study the boiling heat transfer, and great achievements [2, 3] have been obtained. However, since the phenomenon is highly influenced by the micro environment and easily disturbed by the measuring equipment, it is difficult to study the micro microscopic mechanism with the experimental method. Besides of the experimental method, numerical simulations are also widely applied in the studying of the flow and heat transfer problem. Bubble growth and boiling heat transfer can be simulated by coupling the multi-phase flow models [4-8] with the energy equations [9, 10]. Among these multi-phase models, the lattice Boltzmann (LB) method [11] based models are most promising. Due to its kinetic nature, a well-developed theoretical basis, and the ability of self-capturing the interface, LB method has many advantages when simulating multiphase problems. Several LB multiphase models have thus been developed [12-17]. Attempts have also been made to study the phase change problems with these lattice Boltzmann multiphase models [10, 18, 19]. Among them, the single component pseudo-potential LB multi-phase model based phase change models are most widely applied due to their unique advantages such as simplicity and automatic nucleation in boiling simulation. Zhang and Chen [10] firstly proposed the phase change model by coupling the energy equation and the pseudo-potential model. Hazi and Markus [20] improved the energy equation and introduced the Lattice Boltzmann thermal model to solve the equation. This model was soon further developed and applied in different phase change problems [21-26]. In these simulations, the LB methods are adopted to solve both of multiphase flow and energy equations. However, Li et al.'s study [27, 28] shows that the LB thermal models cause errors when applied with the pseudo-potential model. To avoid these errors, they proposed a hybrid LB phase change model by applying a finite difference scheme in the simulation of the energy equation, and obtained the pool boiling curve with the proposed model [29]. This model was then adopted in the research of boiling performance on different heating surfaces [30-33]. Besides of the error terms in the LB thermal model, the conjugated condition treatment is also important in the simulation of boiling phenomenon when the thermal parameter jumps (conjugated heat) exist [34]. In the recent research, special treatments were adopted in the LB thermal equation to solve the conjugated heat problem on the heating wall [35-37]. However, since the thermal parameters across the vapor-liquid interface also changes dramatically in very limited lattices, which can be considered as a series of continuous jumps, the conjugated heat should also be considered to maintain the energy conservation of the system, which has been ignored in most of these works. With regard to the hybrid LB thermal model [29], we also found that the conjugated heat cannot be properly treated with the adopted finite difference schemes, which may lead to numerical errors and instability in the simulation. To solve the problem, in the present work, we introduced an infinite volume discrete scheme to deal with the diffusion term in the energy equation and verified this model with energy conservation law in the simulation of one dimensional heat conduction across the liquid-vapor interface and the liquid film evaporation on the heating wall. The improved hybrid thermal LB model is then applied in the simulation of bubble nucleation and boiling heat transfer.

The rest of this paper is arranged as follows: Section 2 briefly introduces the basic theory of the hybrid pseudo-potential LB model and the corresponding discrete methods of the energy equation; verification of the improved hybrid LB model is carried out in Section 3; the improved hybrid LB model is then applied in the simulation of bubble nucleation and boiling heat transfer in Section 4; finally, a brief conclusion is made in Section 5.

## 2. The hybrid thermal LB model

The hybrid thermal LB model consists of two numerical methods: one is the pseudo-potential model for the simulation of multiphase flow field; the other is the conventional numerical model for energy equation. In this section, the hybrid thermal lattice Boltzmann model proposed by Li et. al. [29] will be briefly introduced, the energy equation and the corresponding finite difference discrete scheme will then be discussed in detail.

### 2.1 MRT pseudopotential lattice Boltzmann method

Through the LB equation of the pseudopotential model is first proposed in the form the Bhatnagar–Gross–Krook (BGK) collision operator [17], the LB equation with a Multi-Relaxation-Time (MRT) collision operator has been proved to be better than the LBGK model in stability and accuracy when the pseudo-potential model is applied [38-40]. The LB equation with a MRT collision operator is given as follows:

$$f_i(\mathbf{x}+\mathbf{e}_\alpha \delta t, t+\delta t) = f_\alpha(\mathbf{x},t) - \left(\mathbf{M}^{-1}\mathbf{\Lambda}\mathbf{M}\right)_{\alpha\beta}(f_\alpha(\mathbf{x},t) - f_\alpha^{eq}(\mathbf{x},t)) + \delta_t \mathbf{F}'_\alpha, \quad (1)$$

where $f_\alpha(\mathbf{x},t)$ is the mass distribution function of particles at node $\mathbf{x}$, time $t$; $\mathbf{e}_\alpha$ is the velocity where $\alpha = 0, 1, 2 \cdots N$; $f_\alpha^{eq}(\mathbf{x},t)$ is the equilibrium distribution. $\mathbf{F}_\alpha$ is the force term in the velocity space. The right side of the equation is a collision operator, and $\mathbf{M}^{-1}\mathbf{\Lambda}\mathbf{M}$ is the collision matrix, in which $\mathbf{M}$ is the orthogonal transformation matrix. In practice, Eq. (1) is solved in two steps as the so called "collision" and "streaming". The collision refers to the right-hand side of Eq. (1) which is the local evolution of the distribution function at time $t$, and the streaming refers to the migration of distribution function from position $\mathbf{x}$ to its adjacent lattices. Using the transformation matrix $\mathbf{M}$, Eq. (1) can also be written as:

$$f_\alpha(\mathbf{x}+\mathbf{e}_\alpha \delta t, t+\delta t) = \mathbf{M}^{-1}\left[\mathbf{m} - \mathbf{\Lambda}(\mathbf{m}-\mathbf{m}^{eq}) + \delta_t(\mathbf{I}-\mathbf{\Lambda}/2)\mathbf{S}\right], \quad (2)$$

where $\mathbf{m} = \mathbf{M}\mathbf{f}$ and $\mathbf{m}^{eq} = \mathbf{M}\mathbf{f}^{eq}$, $\mathbf{I}$ is the unit tensor, and $\mathbf{S}$ is the forcing term in the space with $(\mathbf{I}-\mathbf{\Lambda}/2)\mathbf{S} = \mathbf{M}\mathbf{F}'$. For D2Q9 lattice, the corresponding diagonal matrix $\mathbf{\Lambda}$ is given by (D2Q9)

$$\mathbf{\Lambda} = diag(\tau_\rho^{-1}, \tau_e^{-1}, \tau_\xi^{-1}, \tau_j^{-1}, \tau_q^{-1}, \tau_j^{-1}, \tau_q^{-1}, \tau_\upsilon^{-1}, \tau_\upsilon^{-1},). \quad (3)$$

where $\tau$ represents the relax time, $\tau_\upsilon^{-1}$ is related to the viscosity of the fluid $\upsilon$, the relationship between the relax time and the viscosity is given by $\upsilon = \frac{1}{3}\left(\tau_\upsilon - \frac{1}{2}\right)\delta t$.

The equilibria $\mathbf{m}^{eq}$ is given by [40]:

$$\mathbf{m}^{eq} = \rho\left(1, -2+3|u|^2, 1-3|u|^2, u_x, -u_x, u_y, -u_y, u_x^2-u_y^2, u_x u_y\right) \quad (4)$$

where $\mathbf{u}$ is the macroscopic velocity and $|u| = \sqrt{u_x^2 + u_y^2}$. The macroscopic density and velocity are calculated via

$$\rho = \sum_\alpha f_\alpha(\mathbf{x},t), \quad \rho\mathbf{u} = \sum_\alpha \mathbf{e}_\alpha f_\alpha(\mathbf{x},t) + \frac{\delta t}{2}\mathbf{F}, \quad (5)$$

where $\mathbf{F} = (F_x, F_y)$ is the force acting on the fluid nodes. In the pseudo-potential model, phase separations are simulated by the introduction of interaction force [15, 16], which can be written as:

$$\mathbf{F}_m(\mathbf{x},t) = G\psi(\mathbf{x},t)\sum_{i=0}^{8}\omega_i\psi(\mathbf{x}+\mathbf{e}_i,t)\mathbf{e}_i. \tag{6}$$

where $G$ is the interaction strength, $\omega(|\mathbf{e}_i|^2)$ are the weights, and $\psi(\mathbf{x},t)$ is the effective density. The weights $\omega(|\mathbf{e}_\alpha|^2)$ are given by $\omega(1) = 1/3$ and $\omega(2) = 1/12$.

In practice, the effective density can be obtained with a non-ideal equation of state (EOS) [41]:

$$p_{EOS} = c_s^2\rho + cG[\psi(\rho)]^2/2. \tag{7}$$

To achieve thermodynamic consistency and improve the stability of the model, the forcing scheme proposed by Li et al. [40] is adopted in this work:

$$\mathbf{S} = \begin{bmatrix} 0 \\ 6(u_xF_x + u_yF_y) + \dfrac{\sigma(F_{m,x}^2 + F_{m,x}^2)}{\psi^2\delta_t(\tau_e - 0.5)} \\ -6(u_xF_x + u_yF_y) - \dfrac{\sigma(F_{m,x}^2 + F_{m,x}^2)}{\psi^2\delta_t(\tau_\xi - 0.5)} \\ F_x \\ -F_x \\ F_y \\ -F_y \\ 2(u_xF_x - u_yF_y) \\ (u_xF_x + u_yF_y) \end{bmatrix}. \tag{8}$$

Besides of the interaction force within the fluid, the total force always contains two other forces. One is the buoyant force caused by the density difference, which can be written as

$$\mathbf{F}_b = (\rho - \rho_{ave})\mathbf{g}, \tag{9}$$

where $\rho_{ave}$ is given by the average density in the computational domain, which is widely used in the former research [29, 30, 36].

The other one is the fluid-wall interaction force used to define the wetting condition, which can be written as [42]

$$\mathbf{F}_{ads} = G_w \psi(\mathbf{x}) \sum_{\alpha} \omega_\alpha S(\mathbf{x}+\mathbf{e}_\alpha) \mathbf{e}_\alpha. \tag{10}$$

where, $G_w$ reflects the interaction strength which can be tuned to obtain different contact angles, $\omega_\alpha = w_\alpha/3$, $S(\mathbf{x}+\mathbf{e}_\alpha)$ is given by $\psi(\mathbf{x})s(\mathbf{x}+\mathbf{e}_\alpha)$, in which $s(\mathbf{x}+\mathbf{e}_\alpha)$ is an indicator function that equals 0 for fluid and 1 for solid phase.

## 2.2. Energy equation for temperature field

The temperature equation is derived on the local balance law for entropy [20, 43], which is given by (neglecting the viscous heat dissipation)

$$\frac{dQ}{dt} = \rho T \frac{ds}{dt} = \nabla \cdot (\lambda \nabla T), \tag{11}$$

where, $s$ is the entropy, $\lambda$ is the thermal conductivity. According to the thermodynamic relations, the following equations can be obtained:

$$Tds = du + pdv = \left(\frac{\partial u}{\partial T}\right)_v dT + \left(\frac{\partial u}{\partial v}\right)_T dv + pdv, \tag{12}$$

where, $u$ is the internal energy which equals to $c_v T$, $\left(\frac{\partial u}{\partial v}\right)_T dv + pdv$ is the heat source related to the density change. The latent heat for phase change at constant temperature can be obtained by the integral of the last two terms. Considering $\left(\frac{\partial u}{\partial v}\right)_T = T\left(\frac{\partial p}{\partial T}\right)_v - p$ [20], Eq. (12) can be rewritten as

$$\mathrm{T}ds = du + pdv = c_v dT + T\left(\frac{\partial p}{\partial T}\right)_v dv. \tag{13}$$

The corresponding latent heat is given by

$$h_{\mathrm{lg}} = \int_{v_l}^{v_v} \left[T\left(\frac{\partial p}{\partial T}\right)_v\right] dv. \tag{14}$$

Rewriting Eq. (11) with Eq. (13) and (14) in Euler coordinates, we can get the energy equation of the phase change process:

$$\rho c_v \left[\frac{\partial T}{\partial t} + \mathbf{u} \cdot \nabla T\right] = \nabla \cdot (\lambda \nabla T) + \rho d h_{\mathrm{lg}}. \tag{15}$$

It should be noted that in the literature [23], the latent heat is given by $h_{\mathrm{lg}} = \int_{v_l}^{v_v} \left[T\left(\frac{\partial p}{\partial T}\right)_v - p\right] dv + pv_v - pv_l$, which is the same as the presented one, since for the liquid-vapor phase change process at a constant saturation temperature, the thermal dynamic consistency requires $\int_{v_l}^{v_v} pdv = pv_l - pv_v$.

## 2.3. Discrete schemes for the energy equation

There are two main discrete schemes have been applied in solving the energy equation. Házi et al. [26] introduced the thermal LB equation to solve the energy equation by rewritten Eq. (15) as:

$$\frac{\partial T}{\partial t} + \nabla \cdot (\mathbf{u}T) = \nabla \cdot (k\nabla T) + \frac{1}{\rho c_v} \nabla \cdot (k\nabla T) - \nabla \cdot (k\nabla T) + \left[T - \frac{T}{\rho c_v}\left(\frac{\partial p}{\partial T}\right)_v\right] \nabla \cdot \mathbf{u}, \tag{16}$$

where $k$ is the thermal diffusivity which determined by the relaxation time in the thermal LB equation. However, it has been proved that Eq. (16) cannot be precisely obtained without additional treatment [27]. To solve the problem, Li et al. [29] introduced a finite-difference scheme to solve the energy equation. In this scheme, Eq. (15) is rewritten as

$$\frac{\partial T}{\partial t} = -\mathbf{u}\cdot\nabla T + \frac{1}{\rho c_v}\nabla\cdot(\lambda\nabla T) - \frac{T}{\rho c_v}\left(\frac{\partial p}{\partial T}\right)_v \nabla\cdot\mathbf{u}. \tag{17}$$

For simplicity, we use $K(T)$ to represent the right-hand side of Eq.(17) The classical fourth-order Runge–Kutta scheme [3, 44] is adopted for time discretization

$$T^{t+\delta t} = T^t + \frac{\delta t}{6}(h_1 + 2h_2 + 2h_3 + h_4), \tag{18}$$

where, $h_1 = K(T^t)$, $h_2 = K(T^t + \frac{\delta_t}{2}h_1)$, $h_3 = K(T^t + \frac{\delta_t}{2}h_2)$, $h_4 = K(T^t + \delta_t h_3)$.

The isotropic central scheme proposed by Lee and Lin [45] is adopted for the spatial discretization in the model which can be written as (D2Q9):

$$\left.\frac{\partial\varphi}{\partial x}\right|_{(m,n)} = \left[\varphi_{(m+1,n)} - \varphi_{(m-1,n)}\right]/3 +$$

$$\left[\varphi_{(m+1,n+1)} - \varphi_{(m-1,n-1)}\right]/12 + \left[\varphi_{(m+1,n-1)} - \varphi_{(m-1,n+1)}\right]/12,$$

$$\left.\frac{\partial\varphi}{\partial y}\right|_{(m,n)} = \left[\varphi_{(m,n+1)} - \varphi_{(m,n-1)}\right]/3 +$$

$$\left[\varphi_{(m+1,n+1)} - \varphi_{(m-1,n-1)}\right]/12 + \left[\varphi_{(m-1,n+1)} - \varphi_{(m+1,n-1)}\right]/12, \tag{19}$$

$$\left.\frac{\partial^2\varphi}{\partial y^2}\right|_{(m,n)} = [\varphi_{(m+1,n+1)} + \varphi_{(m-1,n+1)} + \varphi_{(m+1,n-1)} + \varphi_{(m-1,n-1)} +$$
$$4\varphi_{(m+1,n)} + 4\varphi_{(m-1,n)} + 4\varphi_{(m,n+1)} + 4\varphi_{(m,n-1)} - 20\varphi_{(m,n)}]/6 \tag{20}$$

where m, n donate the lattice positions in the two coordinate directions, respectively. It should be noted that the discretion of the term of $\nabla\cdot(\lambda\nabla T)$ with the above scheme requires two neighboring layers of nodes, which may cause difficulties in dealing the boundary condition. Hence, the diffusion term in Eq. (17) is always rewritten as

$$\nabla\cdot(\lambda\nabla T) = \Delta T + (\nabla\lambda)\cdot(\nabla T) \tag{21}$$

before been discretized [27]. Compared with the thermal LB scheme, the finite difference scheme directly solves the energy equation, which is free of error terms. However, with the treatment of Eq. (21), the diffusion term is discretized in the non-conservation form. It can be easily proving that the energy conservation of the entire simulation domain cannot be ensured when the isotropic central scheme is adopted in the diffusion terms. Since the thermal parameters change dramatically across the liquid-vapor interface, large numerical error may be generated at the interface. This problem can be more serious when the solid-fluid interface exists in the simulation domain. The gradient of the thermal conductivity loses its mathematical rigor as the width of the interface is 0 in this case. To maintain the energy conservation, in our previous work [4], a 1th order discretion scheme is applied in the term of $(\nabla \lambda) \cdot (\nabla T)$ to deal with the conjugate heat caused by the $\nabla \lambda$. However, this treatment reduces the accuracy of the model, and the mathematical rigor problem still exists.

To avoid these problems, in this work, the finite volume method [46] is introduced to discretize the diffusion term. As shown in Fig. 1, the control volumes of each lattice are bounded by the middle point of the lattice edges. The heat conducting to the control volume of point P can be written as

$$q_{diff,P} = \int_{\delta_V} \nabla \cdot (\lambda \nabla T) dV, \tag{22}$$

where $\delta_V$ is the control volume of lattice point P. The corresponding discrete form of thermal diffusion term at point P is then given by

$$\nabla \cdot (\lambda \nabla T)\big|_P = \frac{q_{diff,P}}{\delta_V} = \frac{q_{diff,P}}{\delta_x^2}. \tag{23}$$

Utilizing the Gauss theorem, the volume integral in Eq. (22) can be replaced by the surface integrals as follows

$$\int_{\delta V} \nabla \cdot (\lambda \nabla T) dV = \delta_x \left[ \left( \lambda_e \frac{\partial T}{\partial x} \right)_e - \left( \lambda_w \frac{\partial T}{\partial x} \right)_w + \left( \lambda_n \frac{\partial T}{\partial y} \right)_n - \left( \lambda_s \frac{\partial T}{\partial y} \right)_s \right]. \tag{24}$$

The heat conductivities in the above equation are the average thermal conductivities of contiguous lattices. To maintain the energy balance, these conductivities are given as follows

$$\lambda_e = \frac{2\delta_x}{\frac{\delta_x}{\lambda_P}+\frac{\delta_x}{\lambda_E}},\quad \lambda_w = \frac{2\delta_x}{\frac{\delta_x}{\lambda_P}+\frac{\delta_x}{\lambda_W}},\quad \lambda_n = \frac{2\delta_x}{\frac{\delta_x}{\lambda_P}+\frac{\delta_x}{\lambda_N}},\quad \lambda_S = \frac{2\delta_x}{\frac{\delta_x}{\lambda_P}+\frac{\delta_x}{\lambda_S}}. \quad (25)$$

The central difference method is adopted to discretize the gradient in Eq. (24) as

$$\left(\frac{\partial T}{\partial x}\right)_e = \frac{T_E - T_P}{\delta_x},\quad \left(\frac{\partial T}{\partial x}\right)_w = \frac{T_P - T_W}{\delta_x},\quad \left(\frac{\partial T}{\partial x}\right)_n = \frac{T_N - T_P}{\delta_x},\quad \left(\frac{\partial T}{\partial x}\right)_S = \frac{T_P - T_S}{\delta_x}. \quad (26)$$

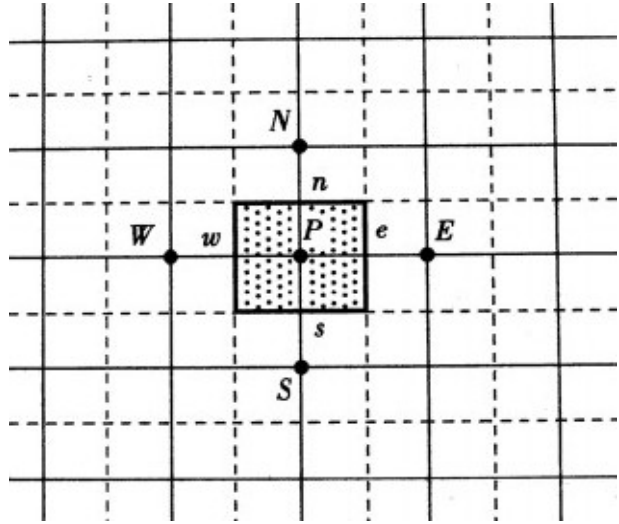

Fig. 1 Schematic of the discrete method.

Besides of the energy conservation, the Runge–Kutta scheme suffer numerical stability problem when the thermal diffusivity is large, since it is essentially an explicit difference scheme. This problem always gets more obvious when the heat conduction within the wall are considered since the conductivity of the wall is much higher than the fluids. In general, an implicit method for the time discretization can avoid this problem. However, the energy equation of the model contains a nonlinear latent heat source term, which is therefore difficult to deduce the simultaneous equations with the implicit method. In the present work, to improve the stability of the model, the discretized time step of the energy equation is set smaller than the time step in the LB flow equation:

$$\delta_{t,T} = \frac{\delta_t}{N}, \tag{27}$$

where, N is a positive integer which can be chosen based on the stability of the model in practice.

In summary, Eq. (25) and (27) are the keys to maintain the energy conservation and improve the stability of the model.

## 3. Numerical validation of the proposed model

In this section, numerical simulations are carried out to validate the performance of the proposed discrete schemes in the simulation of liquid-vapor phase change. For comparison, the isotropic central scheme is also adopted in the simulations.

### 3.1 Non-ideal equation of state and parameters setup

In this work, the widely used Peng-Robinson (P-R) equation of state [41] was adopted to calculate the interaction force:

$$p_{EOS} = \frac{\rho RT}{1-b\rho} - \frac{a\varphi(T)\rho^2}{1+2b\rho-b^2\rho^2}, \tag{28}$$

where $\varphi(T) = \left[1+\left(0.37464+1.54226\omega-0.26992\omega^2\right)\sqrt{1-T/T_c}\right]^2$, $a = 0.45724 R^2 T_c^2 / p_c$ and $b = 0.0778 R T_c / p_c$. Based on the previous research [29], we choose $a = 3/49$, $b = 2/21$ and $R = 1$. The parameter $\sigma$ in Eq. (8) is set to 0.107 for the sake of stability and achieving thermodynamic consistency. The equilibrium state of two-phase coexistence is determined by the saturation temperature $T_s$. In this work, $T_s$ is set as $0.86 T_c$ which is widely used in the previous researches[27, 30-33, 36], and the corresponding coexistence densities are $\rho_l \approx 6.5$ and $\rho_v \approx 0.38$. The kinematic viscosities of the liquid and vapor phases are taken as $\upsilon_l \approx 0.1$ and $\upsilon_v \approx 0.5/3$, respectively. The thermal capacity $c_v$ is set to a constant as 6.0 and the corresponding thermal conductivity is given by $\lambda = \rho c_v k$, where the thermal conductivity $k$ is also set to a constant as 0.1. For simplicity, the lattice parameters are given by $c = \delta_x / \delta_t = 1$ and $\delta_x = \delta_t = 1$.

### 3.2 Heat transition through static phase interface

To show the influence of the discrete schemes of the diffusion term on the performance of the model, we first simulated the heat transfer through a static phase interface. In this case, the fluid is stagnant, hence the influence of the convection terms and latent heat can be eliminated. The simulation domain is chosen as $N_x \times N_y = 120 \times 20$ as shown in Fig. 2. To avoid the density changes during the heat transfer process, the temperature in the EOS is set as a constant of 0.86 $T_c$, hence, the heat transfer doesn't influence the distribution of the density. Initially, the fluid densities for each half of the domain are set as the coexistence densities, respectively, and the temperature within the domain is set as $\frac{(T_h + T_l)}{2}$, where $T_h = 1$ and $T_l = 0$. Steady state temperature distribution within the domain can be obtained after the simulation get convergence. Fig. 3 shows the simulation results of the temperature distribution along the x direction. It can be seen from this figure that the temperature distributions obtained by different discrete schemes are a little different, which shows that the discrete schemes have influence on the heat transfer simulation. To further quantify the difference of these schemes, we compared the relative heat flux difference, which is given by $\frac{2|q_h - q_l|}{(q_h + q_l)}$, between the constant temperature boundaries. It is found that the relative heat flux difference obtained by the infinite volume scheme is under $10^{-10}$, while the relative heat flux difference given by the isotropic central scheme is about 0.18. We further investigated the relative heat flux difference at different temperature of the boundaries of $\Delta T=2,3,4,\ldots,8$, similar results can be obtained. According to the energy conservation law, the heat flux difference at these boundaries should be 0 for a steady state heat transfer, a conclusion can be drawn that the proposed discrete scheme performance much better in energy conservation than the isotropic central schemes.

It should be noted that, in the previous works [35-37, 47] which adopted the thermal LB models to solve the energy equation, the source term of 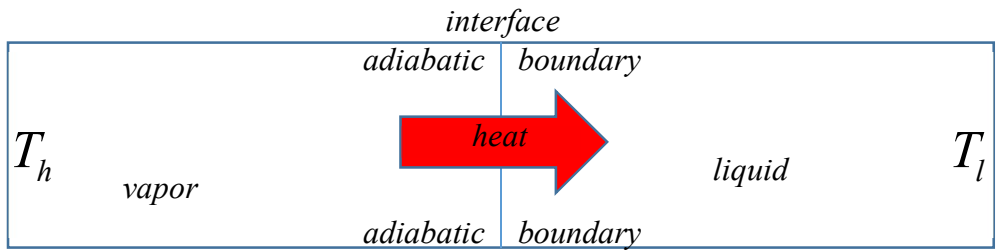 $-\left(\nabla \frac{1}{\rho c_v}\right)\cdot(\nabla T)$ is always introduced to correct the diffusion term obtained by the LB thermal model, which can also lead to energy non-conservation across the interface. Hence, in the multi-phase flow heat transfer problem, the conjugate heat should be considered in liquid-vapor interface as well as the solid-fluid interface.

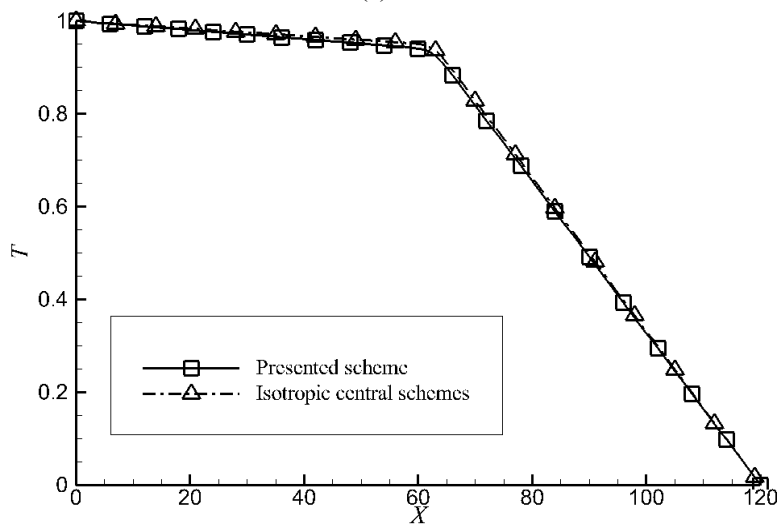

(a)

(b)

Fig. 2 Heat transition through static phase interface: (a) Schematic of computational domain. (b) Simulation results of temperture distribution along the x direction.

## 3.3 Film evaporation on the heating wall

Besides of the phase interface, there is also a thermal conductivity saltation at the fluid-solid boundary when the temperature in both fluid and solid are simultaneously solved (conjugated). To verify the energy conservation of the present model, the film evaporation on the heating surface simulated. As shown in Fig. 1, the simulation domain is chosen as $N_x \times N_y = 20 \times 120$. A static wall with thickness of 20 is placed on the bottom of the simulation domain. The thermal properties are given by $\rho_{\text{wall}} = 3\rho_l$, and $c_{v,\text{wall}} = c_{v,l}$. Initially, a liquid film with thickness of 50 is placed on the bottom of the fluid domain. The bottom boundary is set as a solid wall and a heat flux density of $q$ is applied on the boundary, the half bounce back boundary condition is adopted to deal with the evolution of density distribution function. The open boundary condition of constant pressure is employed at the upper boundary where the density is set constant as $\rho_v$. For a given heat flux, the stream mass flow should satisfy the following equation when the flow gets stable:

$$m'' = q / h_{fg}, \tag{29}$$

where $h_{fg}$ is the latent heat of the phase change at 0.86 $T_c$. It can be calculated with Eq. (14) as $h_{fg} \approx 0.58$. To illustrate the influence of the thermal conductivity on the models, different thermal conductivity ratios are adopted in the simulation. The simulation results are shown Fig. 4.

Fig. 4 (a) is the simulation results given by the isotropic central scheme, it can be seen from these figures that the simulated mass fluxes agree well with the analytical ones when the thermal conductivity ratio is 1. Compared with the simulation results of heat transfer across the static interface, the errors of this scheme seem much smaller, one possible reason is that the temperature gradient at the interface is highly reduced by the evaporation of the fluid compared with the static interface without evaporation. However, the simulation results of the thermal conductivity ratio of 3 given by the isotropic central schemes are obviously larger than the analytical results, showing that unwanted heat source is generated in this case. Further increasing the thermal conductivity ratio, the simulation gets unstable. The simulation results given by the proposed infinite volume model is shown in Fig. 4(b), as we can see in the figure, the mass flux ratios are independent with the thermal conductivity ratios, and agree well with the analytical results. To illustrate the reason of the difference, we further presented the simulated heat flux distribution (Fig. 5) of the evaporation process with different discrete methods. In this case, the heat flux is fixed as 0.0002 with the thermal conductivity ratio of 3. It can be seen from these figures that a heat flux jump appears at the liquid-wall interface, showing that the discontinue energy transfer is generated in this area. While the result given by the infinite volume model show no heat flux difference between the fluid and solid. These results show that the isotropic central schemes adopted in the previous hybrid model cannot maintain the energy balance when the thermal parameter jumps exist in the simulation.

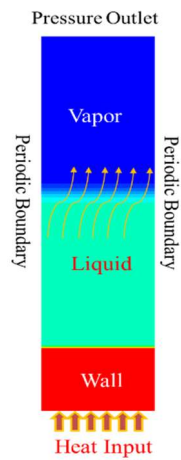

Fig. 3 Schematic of film evaporation on the wall

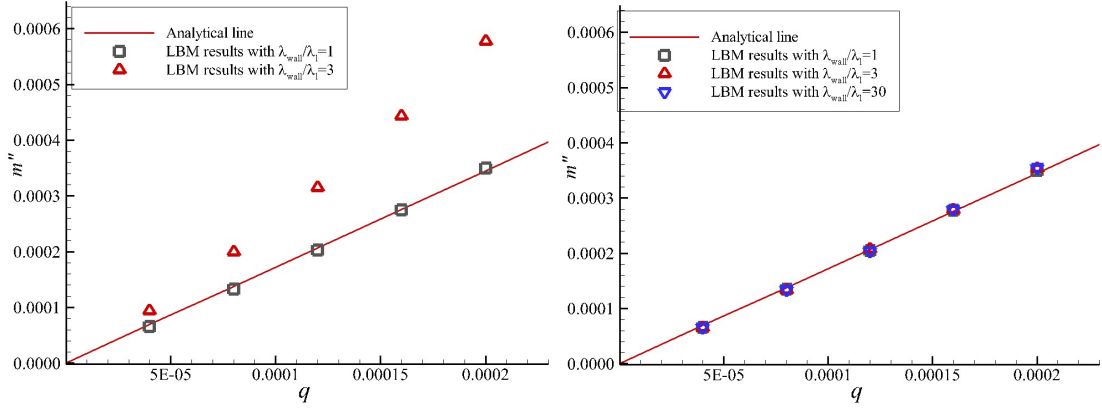

Fig. 4 Comparison of LBM results and analytical results of the thin film evaporation: (a) the isotropic central schemes; (b) the finite volume scheme.

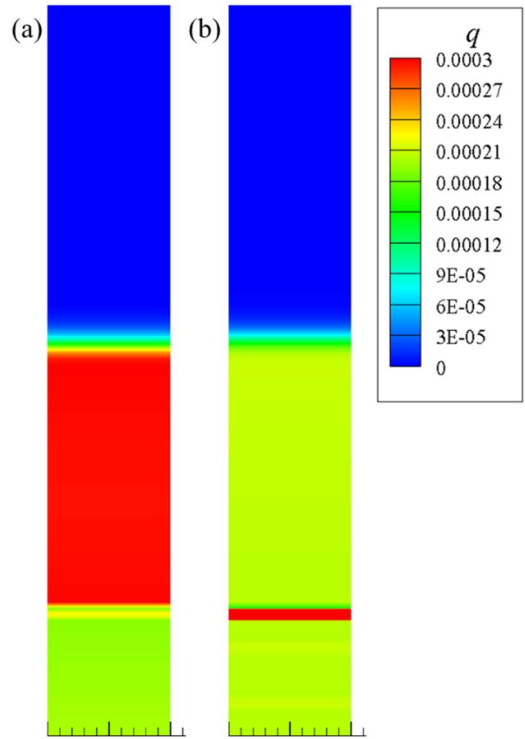

Fig. 5 Heat flux distribution obtained by different discrete schemes: (a) the isotropic central schemes; (b) the finite volume scheme.

## 4 Bubble nucleation and boiling on the hydrophilic wall

In this part, the above-mentioned hybrid thermal lattice Boltzmann model was used to study bubble dynamics and boiling heat transfer on the hydrophilic surface. Boiling curves from the onset of nucleate boiling were obtained numerically. Generally, the bubble nucleation on the even heated wall cannot be a directly simulated by the phase change model without special treatments. To solve the problem, different nucleation formation treatments were proposed. Márkus and Házi [34] initially assigned an existing embryo in their work, which cannot reflect the entire ebullition cycle. Li et al. [31] and Zhang and Chen [35] added small temperature fluctuations in the grid, and successfully simulated the pool boiling phenomenon. However, as pointed by Mu et al.[36], the practice of setting a high temperature point on the heating wall to stimulate nucleation is not necessary, and it is more reasonable and acceptable to set the inherent non-uniformities such as surface wettability and surface roughness as a disturbance to promote the nucleation. Hence, they established a cavity model to study the bubble nucleation on the roughened surface. However, the size of the cavities given in their research is quite large compared to the wall thickness, the heat resistance of the heating wall may become uneven in this case. With a cavity of that size on the wall, the bubble nucleation is still generated by the heating uniformity caused by the geometric change. To our best understanding, the bubble nucleation appears on the flat heating surface where the nucleate sites are much smaller than the bubble, hence the uniformity of the surface should be small enough to eliminate the effect of geometric change of the wall. Besides of the surface roughness of the wall, it is found that the surface wettability also has a great influence on the bubble nucleation [27, 29-31, 35], which provide a possible way to treat the bubble nucleation. Since the treatment of surface wettability does not influence the heat transfer properties of the wall, in this work, a bubble nucleation treatment method is proposed based on the surface wettability adjustment.

## 4.2 Computational setup

As shown in Fig. 6(a), the computational domain is a 2D rectangular with a heating plate located at the bottom wall. The lattice size is setting as 400 × 600. Periodic boundary conditions are imposed on the left and right boundaries. At the top of the computation domain, convective outlet boundary is set. Initially, the computational domain is filled with saturated liquid at $0.86T_c$. Constant temperature boundary condition without any heating fluctuations is applied on the bottom of the wall. The heater size is set as 25 lu. in height. The density and the conductivity of the wall are both set as three times of the liquid, and the heat capacity is set equal to the saturation liquid. Parameters of the fluid is set the same as the previous section. The interaction strength between the solid-liquid $G_w$ is set as 0.3 and the corresponding contact angle is about 35°. As the bubble nucleation requires the fluid to overcome the solid-fluid interaction force, it can be assumed that the interaction strength is smaller at the nucleation site. Based on the idea, the nucleation sites are treated by introducing linear interaction strength drops as shown in Fig. 6 (b). The width of each nucleation site is set as 10 lu., which is about two times of the liquid-vapor interface width. Since in practice, the required activation energies for different nucleation sites are different, the lowest interaction strengths in this work are set different. In this work, the lowest interaction strengths (from left to right) are set as 0, -0.3, -0.6, respectively.

In our simulation, variables are given in a dimensionless form as shown in ref. [30]. The reference length $l_0$, reference velocity $u_0$ and reference time $t_0$ are given by

$$l_0 = \sqrt{\frac{\sigma}{g(\rho_l - \rho_v)}}, \quad u_0 = \sqrt{gl_0}, \quad t_0 = l_0 / u_0 \tag{30}$$

where $\sigma$ is the surface tension, $\rho_l$ and $\rho_v$ are densities of saturated liquid and saturated vapor, respectively. These three characteristic values provide a convenient way to analyze the simulation results as long as a relation between the lattice units and international system of units (ISU). However, since a temperature related parameter is missing among the above reference values, the heating parameters such as the heat flux and temperature difference cannot be nondimensionalized. To solve the problem, a reference heat flux density is introduced in the present work, which is given by

$$q_0 = (\rho_l - \rho_l)h_{lg}. \tag{31}$$

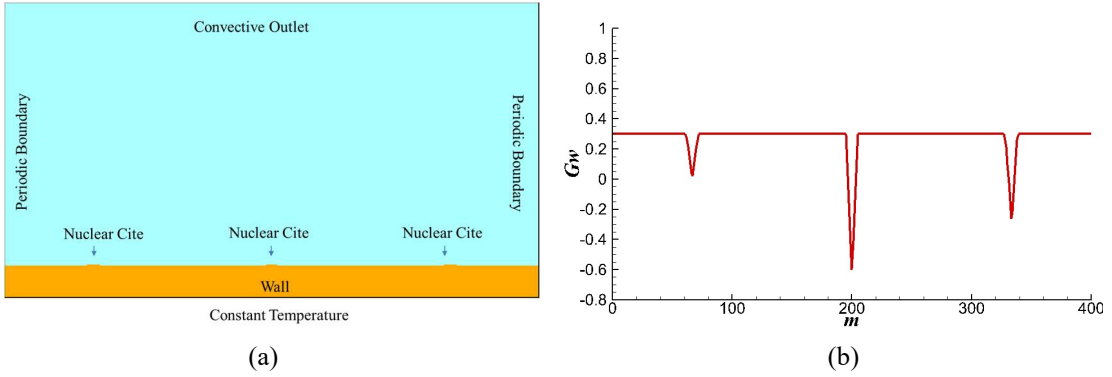

(a)                                                     (b)

Fig. 6 Schematic of computational domain: (a) boundary conditions; (b) nucleation site treatment.

## 4.3 Bubble nucleation

Figures. 7-9 show the bubble nucleation sites activation at different dimensionless superheat Ja, which is defined by $Ja = c_{p,l}(T_w - T_s)/h_{lg}$. According to these simulation results, the number of nucleation sites increases with the superheat of the wall, which has been observed in many experimental research [1, 48]. Bubble nucleation sites with smaller interaction strength active at smaller superheat degree, showing less activation energy are required, which agrees with the former analysis in the work.

Fig. 7 shows the bubble nucleation and growth when $Ja \leq 0.093$. Since the superheat is low in this case, only the bubble nucleation site at the center of the wall is activated. As shown in the Fig. 7 (a), at $t^* = t/t_0 = 47.6$, one bubble departed and the bulk liquid flooded in and contact with the heating surface. Then transient heat conduction from heater to liquid happened, but no bubble emerged at this period (from $t^* = 51.3$ to $t^* = 59.8$). At the end of the waiting period, one bubble occurred and a new cycle began. Similar results can be also found in Fig. 7(b), of which $Ja = 0.114$. The waiting period (from $t^* = 18.3$ to $t^* = 22.0$) also existed in this case, however, it is much reduced compared with the results at $Ja=0.093$. Since the natural convection heat transfer during waiting period is low compared to boiling heat transfer, the heat transfer coefficient increases with the superheat due to the decreasing waiting period [49].

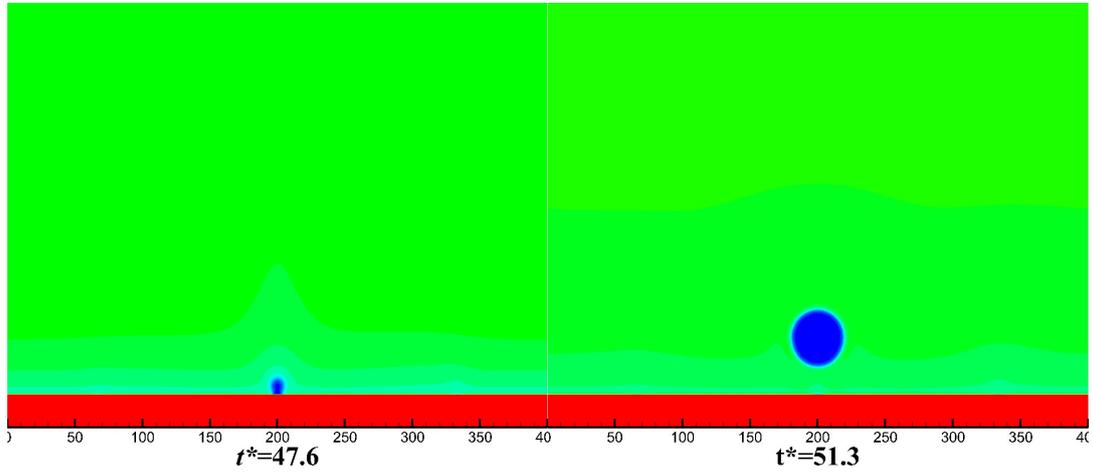

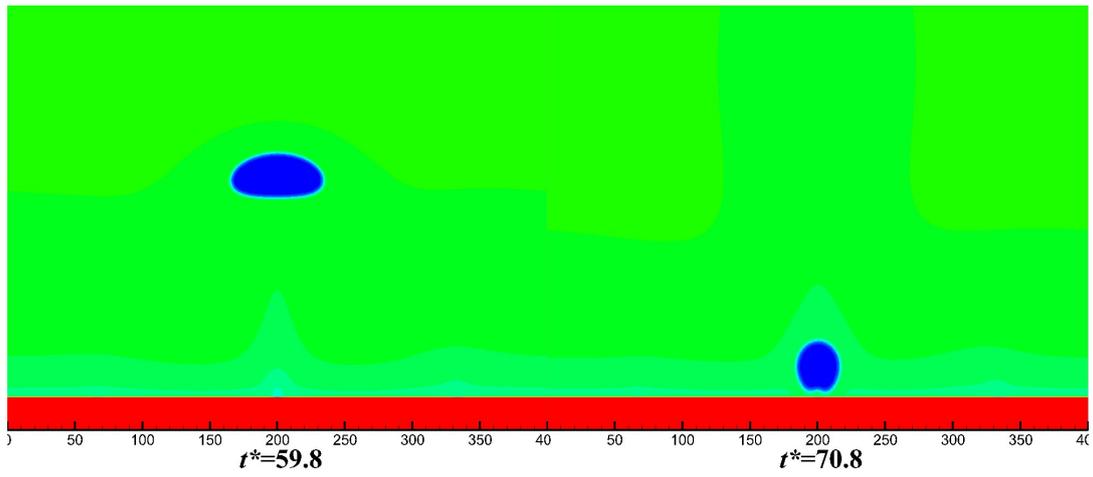

(a)

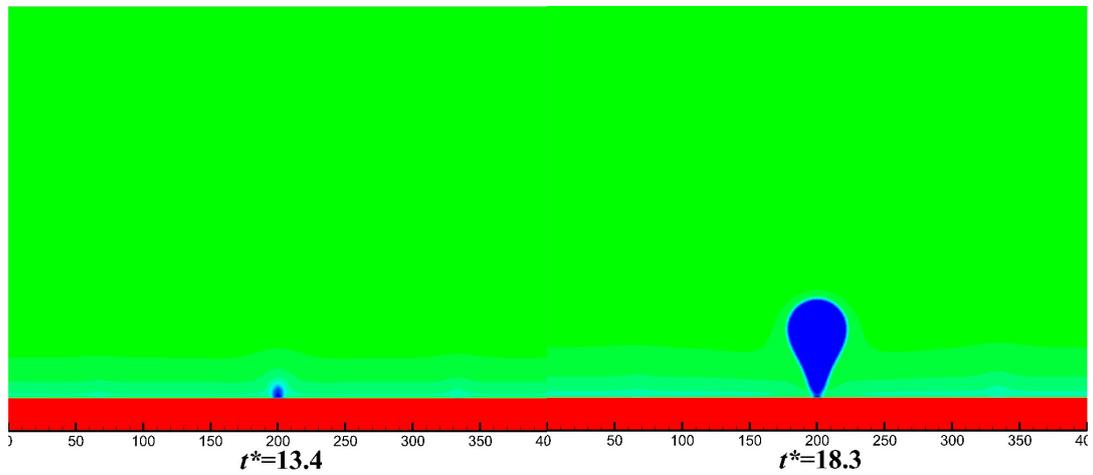

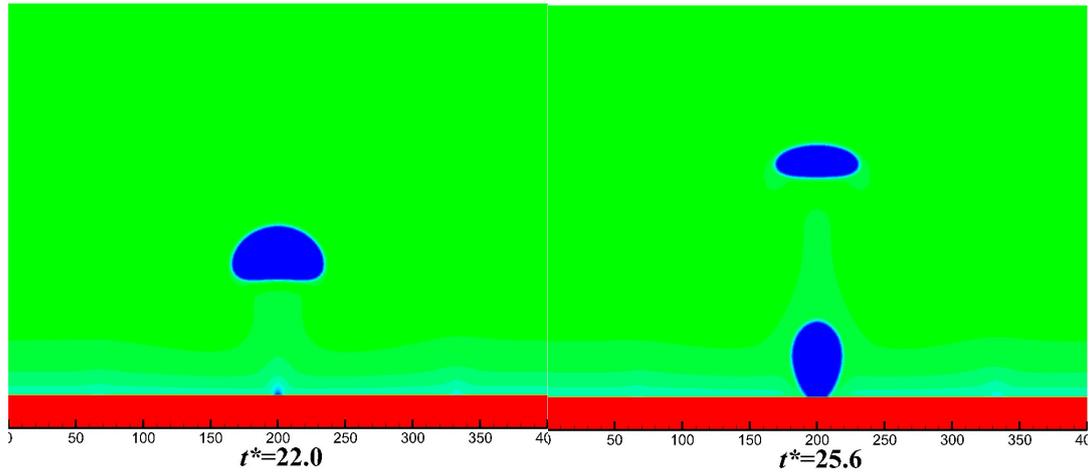

(b)

Fig. 7 Snapshots of the boiling process for a single activited bubble nucleation site: (a) Ja=0.093, (b) Ja=0.114

As the superheat grows, the nucleation site on the right of the wall is also activated (shown in Fig. 8(a). Similar with the first activated bubble, the waiting period also exists during the growing period of the bubble. However, as shown in these figures, the waiting period of the second bubble is much longer than the first one. It can be found that the shape of the bubble at the center of the wall is deformed when the second nucleation site is activated, showing that the interaction effects exist when multi-bubble growing on the heating surface. The third nucleation site activated at $Ja$=0.145, as shown in Fig. 8(b). The boiling heat transfer exists at all these three nucleation sites simultaneously, however, with different departure frequency. It should be noted that, as the superheat increase, the waiting period of the bubble in the middle disappears, a vapor column is consistently maintained on the wall after the bubble departed due to the fast phase change rate.

Continually increasing the superheat of the wall after the preset nucleation sites are all activated. It is found that the bubble nucleation also appears at the surface with uniform wetting condition (Fig. 9(a)). As the superheat grows, more nucleation sites appear at the uniform wetting surface (Fig. 9 (b)), which enhances the heat transfer at the surface. It also can be seen from these figures that, as the number of bubble nucleation sites and the departure frequency increase, the interaction among these bubbles becomes stronger: the raising paths of the departed bubbles are not straight upward, and the shape of the bubbles become less regular, showing flow disturbance is generated by the interactions.

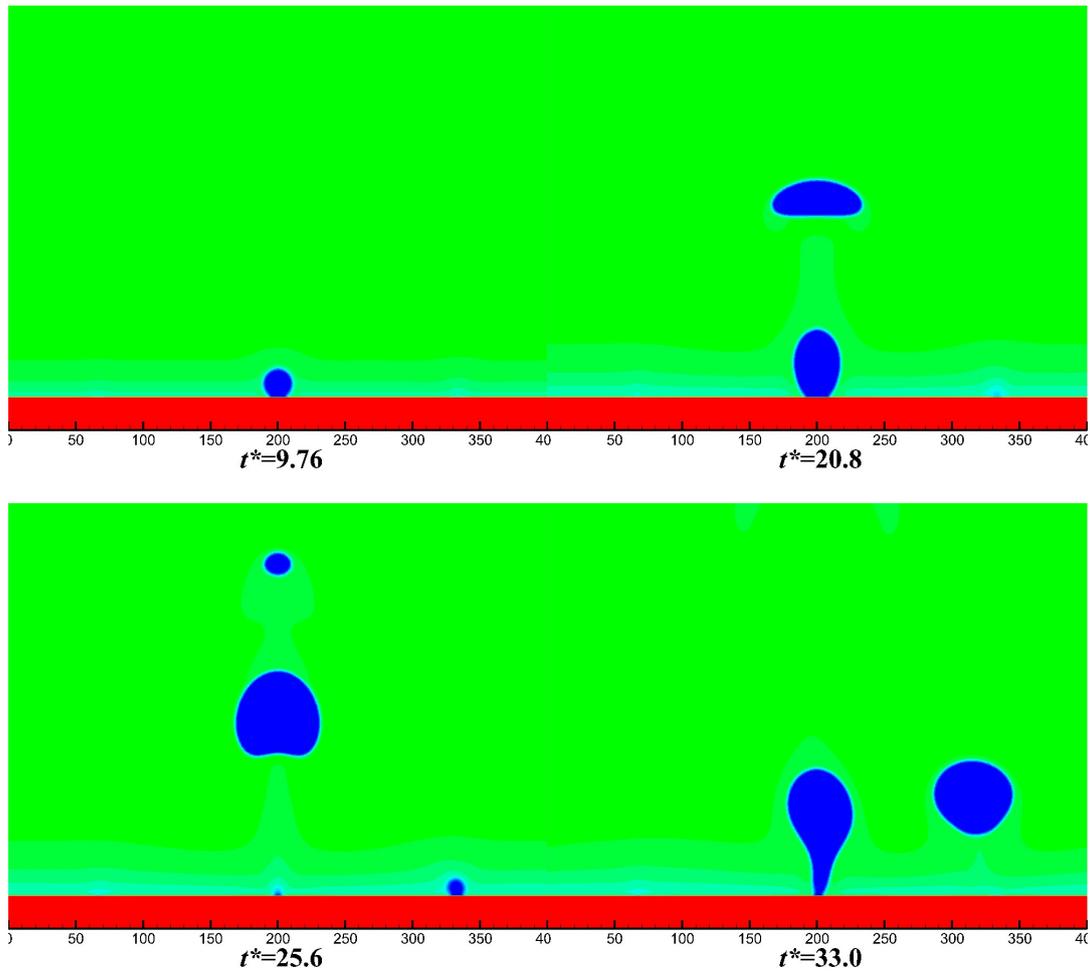

(a)

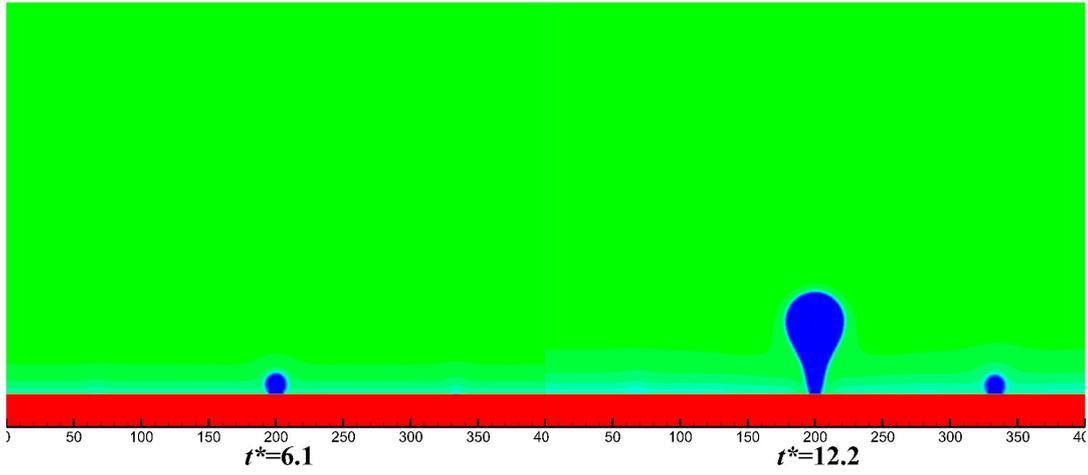

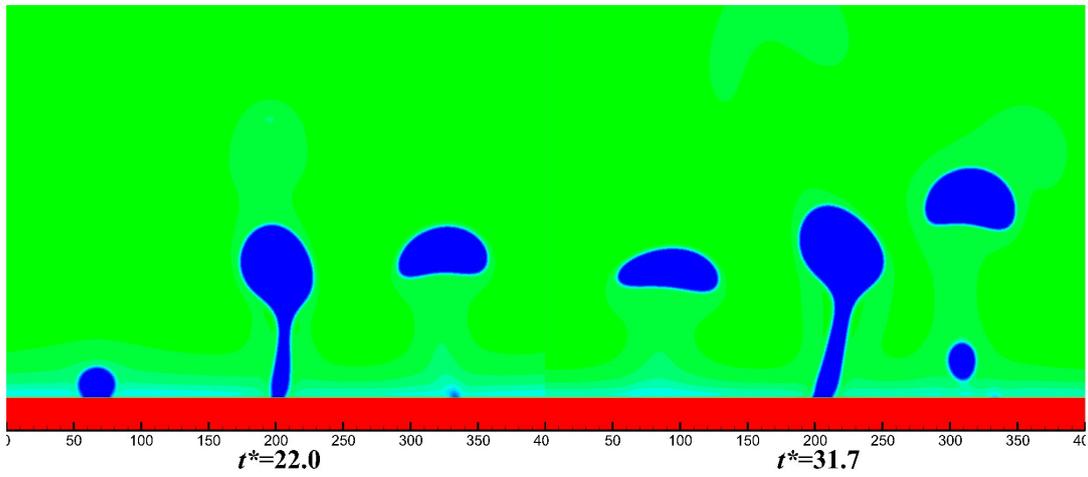

(b)

Fig. 8. Snapshots of the boiling process with multiply activited bubble nucleation sites: (a) Ja=0.124; (b) Ja=0.145.

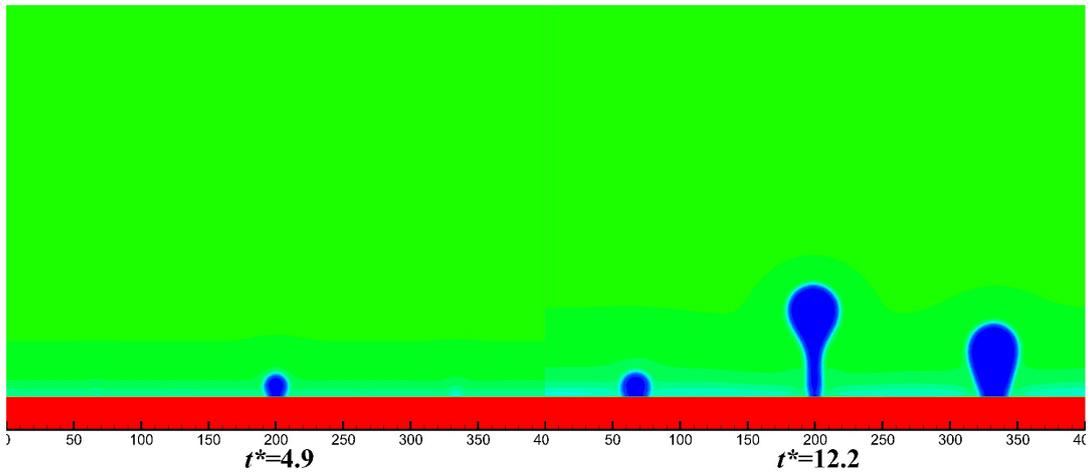
24

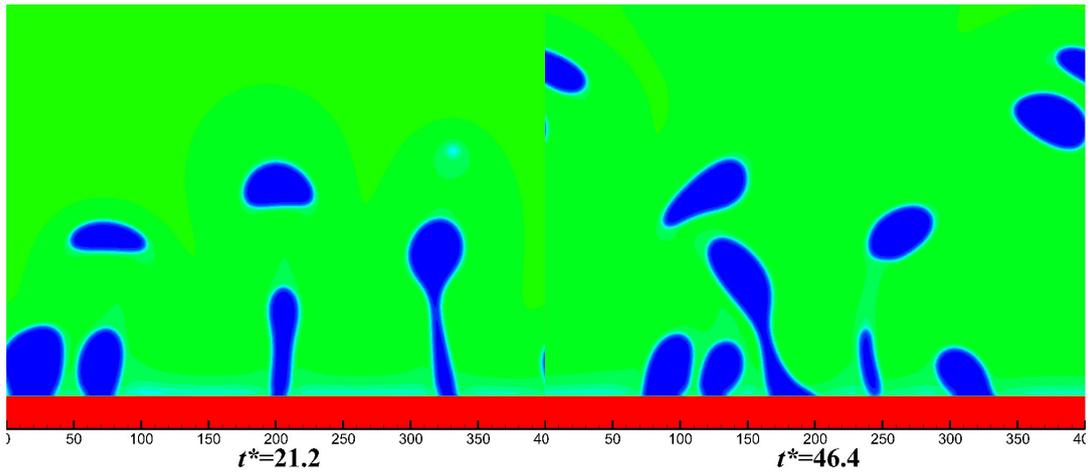

(a)

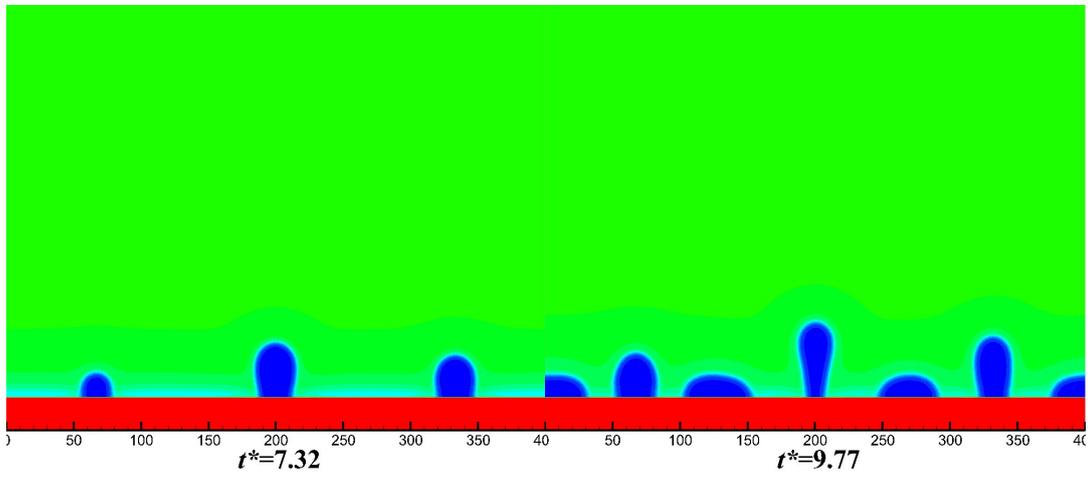

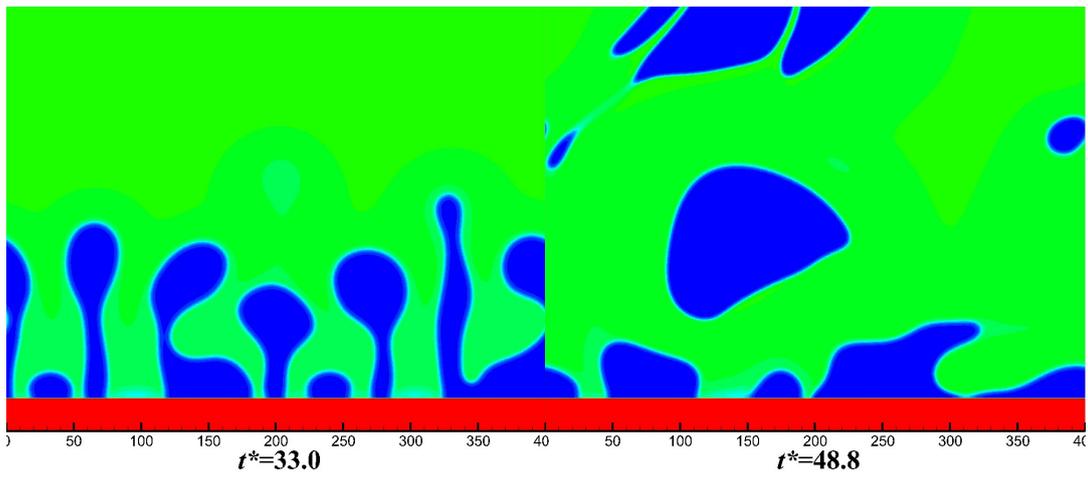

Fig. 9 Snapshots of the boiling process with multiply activited bubble nucleation site: (a) Ja =0.166 (b) Ja=0.197

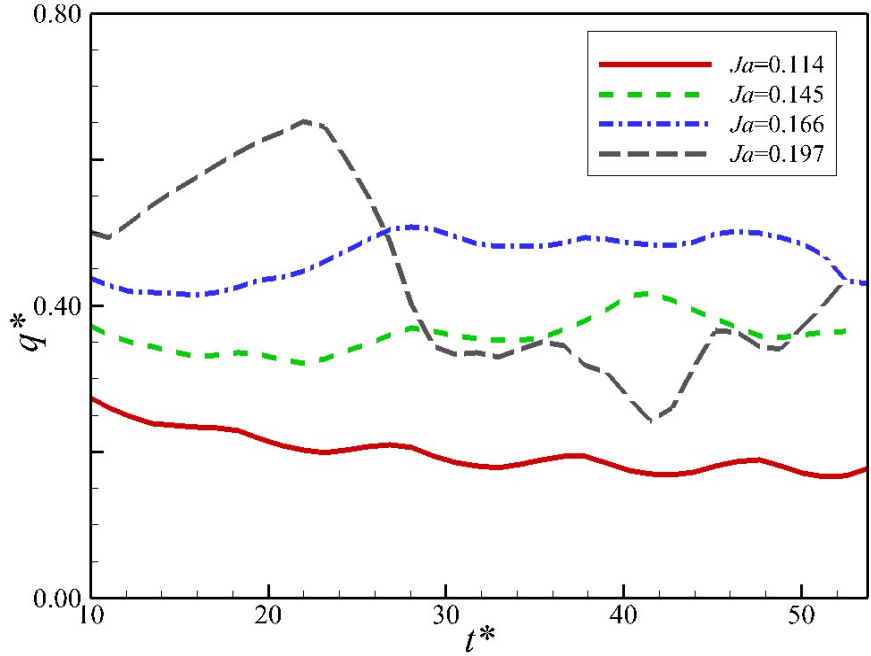

Fig. 10 Transient heat flux for different overheat.

To compare the heat transfer characters at different nucleation condition, we presented the simulated transient heat flux in Fig. 10. The dimensionless heat flux is given by $q^* = \bar{q}/q_0$, where $\bar{q}$ is the average heat flux on the bottom of the wall. It can be seen from this figure that the heat flux increases with the superheat, except the heat flux given by Ja=0.197. The heat flux fluctuation shows an obvious periodicity when only a single nucleation site is activated (Ja=0.114), while this periodicity becomes much less apparent as increase of the superheat since more nucleation sites are activated. When Ja=0.197, the heat flux increases quickly at the early stage of the boiling process due to the activation of large number of nucleation sites. However, a sharp decline of heat flux appears when $t^* \geq 22$, this is mainly due to the bubble coalescence at the wall surface, leading to a decrease of wetting area. Since the heat transfer efficiency of the vapor is much smaller than the liquid, the heat flux is significantly reduced in this case. These results show that the boiling becomes very unstable when the transition boiling emerges since it can be viewed as the combination of the nucleate boiling and film boiling [29, 50].

## 4.4 Boiling curve and boiling regimes

In order to study the influence of the superheat and corresponding bubble dynamics on boiling heat transfer coefficient more clearly, the boiling curve [2, 51] is presented in Fig. 11(a), which is established by averaging the transient heat flux over time. For the cases of $Ja < 0.093$ (onset of nucleate boiling (ONB)), the wall superheat is insufficient to support bubble formation and growth. Hence no boiling occurs and the heat flux is very low as can be seen in Fig. 11(a). With the increase of the wall superheat, boiling will occur and the heat flux will increase rapidly until reaching its maximum value (critical heat flux (CHF)) around $Ja = 0.186$. After that the heat flux decreases dramatically as the bulk liquid will be insulated from the heating wall by more and more vapor. According to the pool boiling theory [2], the pool boiling curve can be divided into four boiling regimes according to the bubble forms during the evaporation. The simulated bubble forms of each regime are given in Fig. 12. When the superheat $Ja$ is at the range of 0.093 to 0.166, the active sites are separated, which refers to the isolated bubble regime. Increasing surface superheat, more and more sites become active, and the bubble departure frequency at each site generally increases. Eventually the active sites are spaced so closely that bubbles merge together during the final stage of growth and release, forming vapor slugs and columns as shown in Fig. 12(b). Corresponding Ja is at the range from 0.166 to 0.187 refer to as the regime of slugs and columns. As the flow rate of the vapor continually increases, the liquid cannot reach the surface fast enough to keep the surface completely wetted with liquid, the heat flux decreases as the wall superheat increase. This regime is referred to the transition regime (Fig. 12 (c)), and the corresponding Ja is at the range from 0.197 to 0.217 in the present work. When Ja is larger than 0.22, the wall surface is fully insulated by vapor, making the transition to the film boiling regime, the heat flux researches its lowest value.

Fig. 11(b) shows the dimensionless heat transfer coefficient (defined by $h^* = q^*/Ja$) at different superheat. It can be seen from this figure that, similar with the heat flux, the heat transfer coefficient increases rapidly after the ONB point, and significantly drops after reaches its maximum value. The corresponding superheat of the maximum heat transfer coefficient is at Ja=0.176~0.186, which is smaller than the heat flux at the CHF [52]. The tendency agrees with the simulated results in previous research, except that the different between the superheat at the maximum heat transfer coefficient and the CHF is smaller than the previous studies [29, 30]. The increase of heat transfer coefficient can be explained by many reasons such as increase of active nucleation sites number and bubble departure frequency. Besides of these factors, we found that during the bubble growing process, the bubbles on the left and right side moved horizontally toward the bubble at the middle of the surface as shown in Fig. 13. The horizontal moving of bubbles may be caused by the natural convection and Marangoni effect [53], as the surface tension of the bubble are uneven due to the temperature gradient. Similar phenomenon has also been perceived in many experimental research [54, 55]. Since the horizontal movement of bubble can increase both of the nucleation site number and bubble departure frequency, the superheat of the maximum heat transfer coefficient is increased.

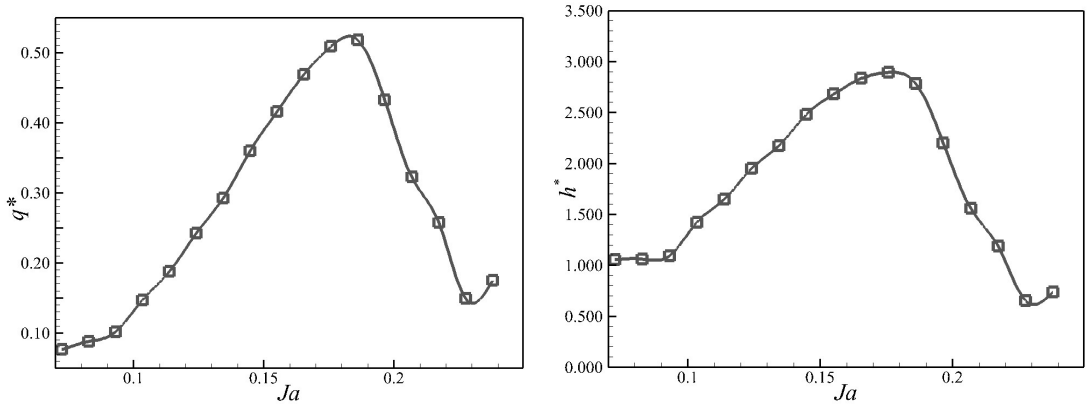

Fig. 11 heat transfer efficiency for different overheat: (a) heat flux; (b) heat transfer coefficient.

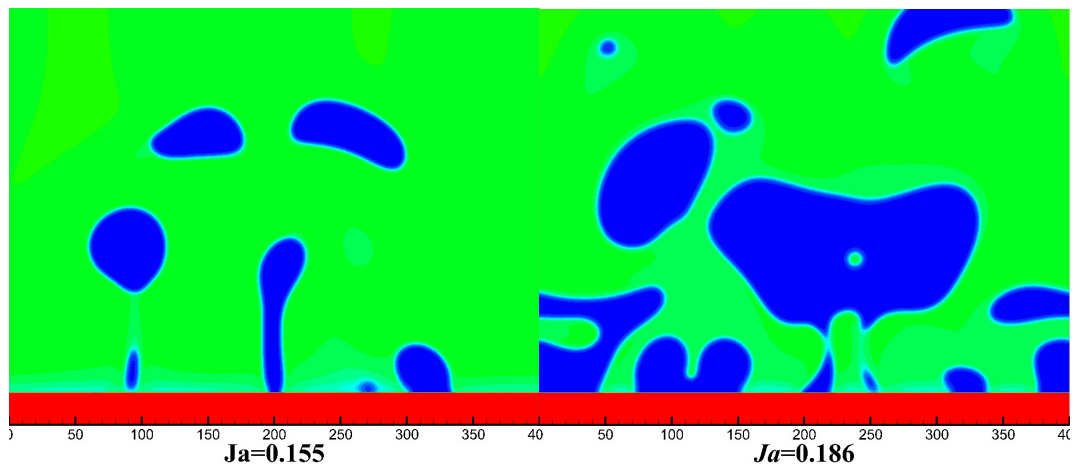

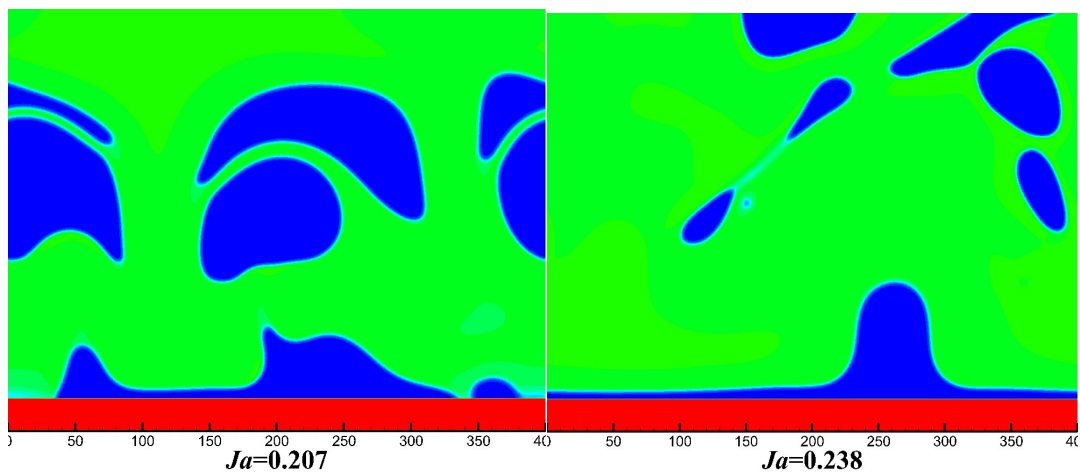

Fig. 12 Snapshots of bubble forms of each boiling regime.

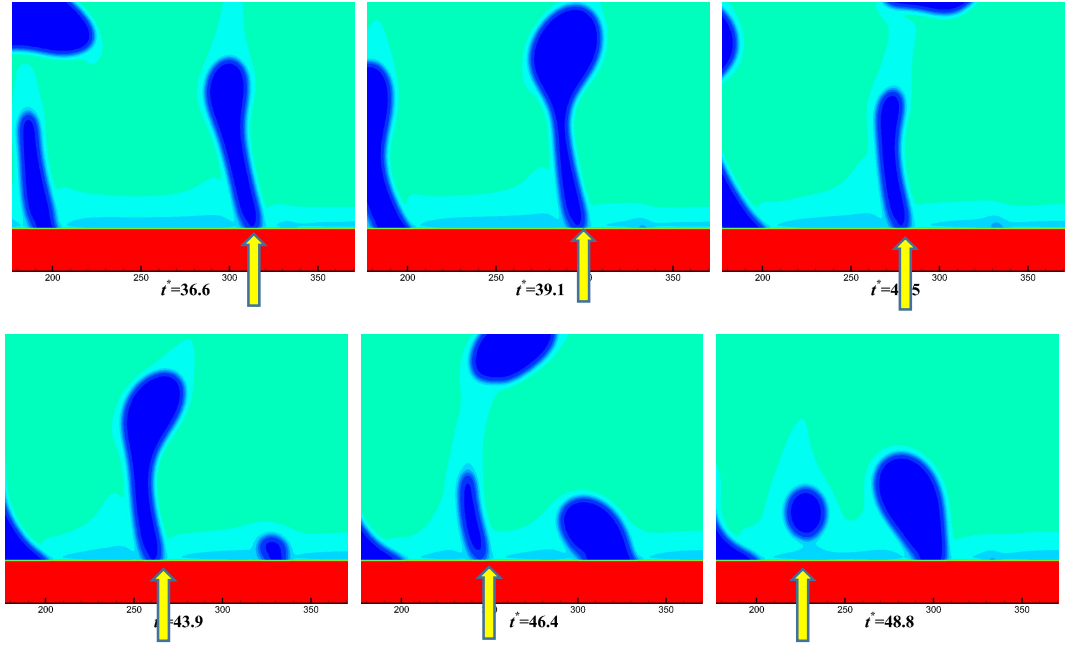

Fig. 13 Snap shots of horizontal moving bubble.

## 5. Summary and conclusion

In this paper, the energy non-conservation problem of the hybrid thermal LB model has been pointed out. To solve this problem, an infinite volume discrete scheme is introduced to deal with the diffusion term in the energy equation. After verified this model with the energy conservation law, the improved hybrid thermal LB model is then applied in the simulation of bubble nucleation and boiling heat transfer. The following findings can be drawn from the present study:

(1) In the simulation of heat transfer between vapor-liquid multi-phase flow and solid wall, the conjugated heat should be considered in both of the vapor-liquid and the solid-fluid interfaces due to the significant heat conductivity change across the interfaces. The isotropic central scheme which widely applied in the hybrid thermal LB model cannot ensure energy conservation in the simulation of heat transfer across the interface, which may lead to numerical errors and instability in the application of the model.

(2) The conjugated heat problem can be solved by dealing the diffusion term with the volume discrete scheme. The stability can also be improved by reducing the time step size in the discretion of energy equation.

(3) With an innovative nucleation site treatment on the wall, the complete boiling stages as well as the boiling curve were successfully reproduced in our simulations without adding any heating fluctuations. The bubble nucleation, growth, departure, and coalescence being well captured. Some basic features of boiling heat transfer were clearly observed in the numerical results, such as bubble nucleation site activation, bubble departure frequency, character of bubble form at different regime, and horizontal movement of bubble during the boiling process.

**Acknowledgments**

This work is sponsored by <National Natural Science Foundation of China> under Grant <51606159>.